\documentclass[wssquare]{ws-procs9x6}            % default, citations in superscript
\usepackage{graphicx,upgreek}
\usepackage[utf8]{inputenc}
\usepackage[T1]{fontenc}
\graphicspath{{../Figures/}}

\begin{document}
\title{The new Felsenkeller 5 MV underground accelerator}

\author{
Daniel Bemmerer$^{1,*}$,
Thomas E. Cowan$^{1,2}$,
Alexander Domula$^{2}$,
Toralf Döring$^{1}$,
Marcel Grieger$^{1,2}$,
Sebastian Hammer$^{1,2}$,
Thomas Hensel$^{1,2}$,
Lisa Hübinger$^{1,2}$,
Arnd R. Junghans$^{1}$,
Felix Ludwig$^{1,2}$,
Stefan E. Müller$^{1}$,
Stefan Reinicke$^{1,2}$,
Bernd Rimarzig$^{1}$,
Konrad Schmidt$^{2}$,
Ronald Schwengner$^{1}$,
Klaus Stöckel$^{1,2}$,
Tamás Szücs$^{1}$,
Steffen Turkat$^{2}$,
Andreas Wagner$^{1}$,
Louis Wagner$^{1,2}$,
Kai Zuber$^{2,\S}$
}

\address{%
$^1$ Helmholtz-Zentrum Dresden-Rossendorf (HZDR), Bautzner Landstr. 400, 01328 Dresden/Germany\\
$^*$ E-mail: d.bemmerer@hzdr.de \\
$^2$ Technische Universität Dresden, Zellescher Weg 19, 01069 Dresden/Germany\\
$^\S$ E-mail: zuber@physik.tu-dresden.de
}

\begin{abstract}
The field of nuclear astrophysics is devoted to the study of the creation of the chemical elements. By nature, it is deeply intertwined with the physics of the Sun. The nuclear reactions of the proton-proton cycle of hydrogen burning, including the $^3$He($\alpha$,$\gamma$)$^7$Be reaction, provide the necessary nuclear energy to prevent the gravitational collapse of the Sun and give rise to the by now well-studied pp, $^7$Be, and $^8$B solar neutrinos. The not yet measured flux of $^{13}$N, $^{15}$O, and $^{17}$F neutrinos from the carbon-nitrogen-oxygen cycle is affected in rate by the $^{14}$N(p,$\gamma$)$^{15}$O reaction and in emission profile by the  $^{12}$C(p,$\gamma$)$^{13}$N reaction. The nucleosynthetic output of the subsequent phase in stellar evolution, helium burning, is controlled by the $^{12}$C($\alpha$,$\gamma$)$^{16}$O reaction. 

In order to properly interpret the existing and upcoming solar neutrino data, precise nuclear physics information is needed. For nuclear reactions between light, stable nuclei, the best available technique are experiments with small ion accelerators in underground, low-background settings. The pioneering work in this regard has been done by the LUNA collaboration at Gran Sasso/Italy, using a 0.4 MV accelerator. 

The present contribution reports on a higher-energy, 5.0 MV, underground accelerator in the Felsenkeller underground site in Dresden/Germany. Results from $\gamma$-ray, neutron, and muon background measurements in the Felsenkeller underground site in Dresden, Germany, show that the background conditions are satisfactory for nuclear astrophysics purposes. The accelerator is in the commissioning phase and will provide intense, up to 50\,$\upmu$A, beams of $^1$H$^+$, $^4$He$^+$, and $^{12}$C$^+$ ions, enabling research on astrophysically relevant nuclear reactions with unprecedented sensitivity.
\end{abstract}


\bodymatter

% =================================================================================
\section{Introduction}
\label{sec:intro}

Driven by the success of recent, large underground neutrino detectors for the $^8$B \cite{SNO13-PRC_all3phases,SuperK11-PRD}, $^7$Be \cite{Borexino11-PRL}, and pp \cite{Borexino14-Nature} neutrinos, a separate but linked field has evolved in recent years: Underground nuclear astrophysics. 

In stars, nuclear reactions take place in a narrow energy window called the Gamow peak \cite{Iliadis15-Book}. Its position is determined by the maximum of a function constructed as the product of two strongly energy (or velocity) dependent functions: First, the Maxwell-Boltzmann distribution of particle energies in the hot stellar plasma, and second, the nuclear reaction cross section. As the relevant energies are below the barrier energy given by Coulomb repulsion, the cross section drops quickly with decreasing energy. 

For zero relative angular momentum between the two reaction partners, the energy-dependent cross section $\sigma(E)$ can be conveniently parameterized by the so-called astrophysical S factor $S(E)$ given by 
\begin{equation}
S(E) = \sigma(E) \times E \times \exp \left[31.29 Z_1 Z_2 \sqrt{\upmu/E} \right] \label{eq:Sfactor}
\end{equation}
with $E$ the center of mass energy in keV, $Z_{1,2}$ the nuclear charge numbers of the two interacting particles, and $\upmu=m_1m_2/(m_1+m_2)$ the reduced mass, in a.m.u., calculated from the masses $m_{1,2}$ of the two interacting particles.

\begin{figure}[tb]
\centering
\includegraphics[width=\textwidth]{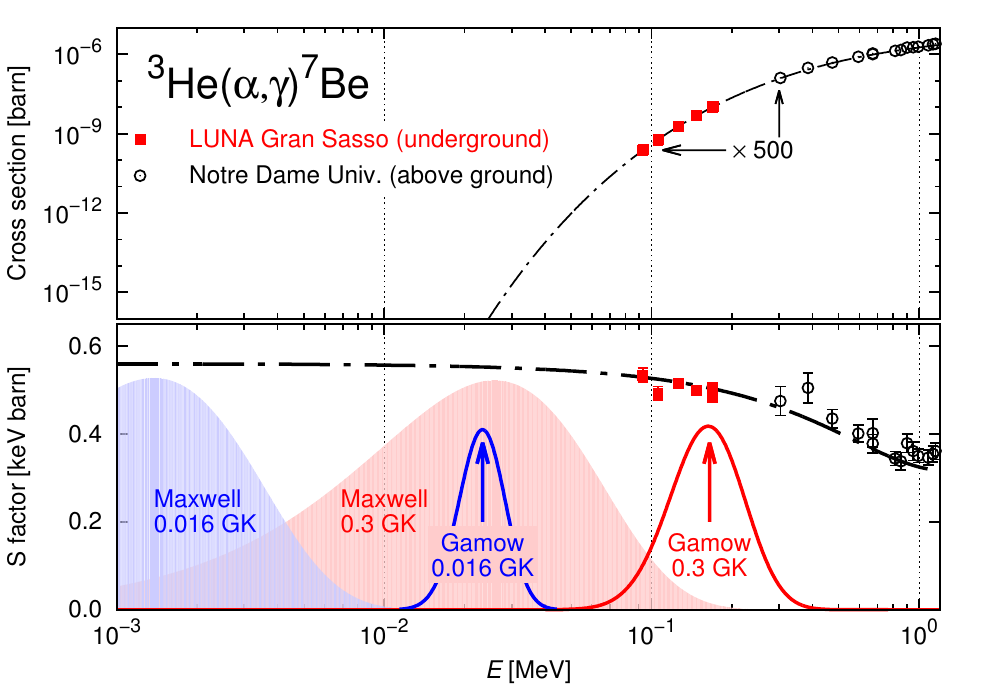}
\caption{Top panel: Experimental cross section data on the $^3$He($\alpha,\gamma$)$^7$Be reaction by the underground experiment LUNA (red filled squares, \cite{Bemmerer06-PRL,Confortola07-PRC}), and by the experiment that reached the lowest energy achievable above ground (black open circles, \cite{Kontos13-PRC}). The lowest cross section measured above ground is 500 times higher than the lowest cross section measured underground, as marked by the arrows. Bottom panel: Cross sections converted to the astrophysical S-factor using equation \ref{eq:Sfactor}. The dot-dashed lines give the recommended cross section by the "Solar Fusion II" group \cite{Adelberger11-RMP}.}
\label{fig:3He4He}       % Give a unique label
\end{figure}

The relevant energies are illustrated in Figure \ref{fig:3He4He}. There, the $^3$He($\alpha,\gamma$)$^7$Be reaction, which controls the passage from the pp-I chain to the pp-II and pp-III chains, is used as an example. For the Sun ($T$ = 0.016 GK), the Maxwell-Boltzmann distribution peaks at 1\,keV, but the Gamow peak for this reaction peaks at approximately 20\,keV. For a typical Big Bang temperature of 0.3 GK, both the Maxwell-Boltzmann distribution and the Gamow peak are shifted to higher energies.

% =================================================================================
% =================================================================================

The steep energy dependence of the cross section, over many orders of magnitude, is removed in the astrophysical S factor in equation (\ref{eq:Sfactor}). 

It is clear from the figure that in order to approach the solar Gamow peak, ultra low background facilities are needed. 
In accelerator-based experiments, such conditions can only be reached by placing the entire laboratory, including the accelerator, in an underground laboratory. 
This is the approach taken with great success by the LUNA (Laboratory for Underground Nuclear Astrophysics) collaboration, at the LUNA 0.4\,MV accelerator deep underground in the Gran Sasso national laboratory, Italy \cite{Broggini18-PPNP}.

For the example of the $^3$He($\alpha,\gamma$)$^7$Be reaction (Figure \ref{fig:3He4He}), cross sections could be measured \cite{Bemmerer06-PRL,Confortola07-PRC} that are 500 times lower than the lowest energy data point taken by any experiment at the Earth's surface \cite{Kontos13-PRC}. 

With the same approach, the $^{14}$N(p,$\gamma$)$^{15}$O reaction controlling the CNO cycle in the Sun \cite{Formicola04-PLB,Lemut06-PLB}, the reactions controlling Big Bang production of $^6$Li \cite{Anders14-PRL,Trezzi17-APP} and $^7$Li \cite{Bemmerer06-PRL}, and several reactions of higher hydrogen burning have recently been studied. 

The most recent example is the $^{22}$Ne(p,$\gamma$)$^{23}$Na reaction, where owing to the low background underground three new resonances have been discovered \cite{Cavanna15-PRL,Bemmerer18-EPL} and two hypothetical resonances ruled out experimentally \cite{Ferraro18-PRL}. These data address the well-known anticorrelation of sodium and oxygen abundances in globular cluster stars \cite{Slemer17-MNRAS}.

% =================================================================================
\section{Scientific motivation}

The limited energy range of the unique LUNA 0.4\,MV underground accelerator has led to a call for new, higher-energy underground accelerators \cite{NuPECC17-LRP,WhitePaper2016}. As a result, new underground accelerator laboratories are under construction or starting up on three continents: The LUNA-MV 3.5\,MV accelerator in Gran Sasso, Italy \cite{Guglielmetti14-DARK}, the CASPAR 1\,MV accelerator in Lead, South Dakota, USA \cite{Robertson16-EPJWOC}, and the JUNA accelerator complex in the Jinping laboratory, China \cite{Liu16-EPJWOC}.

The present contribution reports on the status of the Felsenkeller shallow-underground accelerator for nuclear astrophysics, updating Ref.~\cite{Bemmerer16-arxiv}.

% =================================================================================
\section{Solar fusion and the $^3$He($\alpha,\gamma$)$^7$Be reaction}

The need for new, higher-energy underground accelerators arises firstly from the need to study hydrogen burning reactions over a wide energy range. 

% =================================================================================
\begin{figure}[b]
\centering
\includegraphics[width=\textwidth]{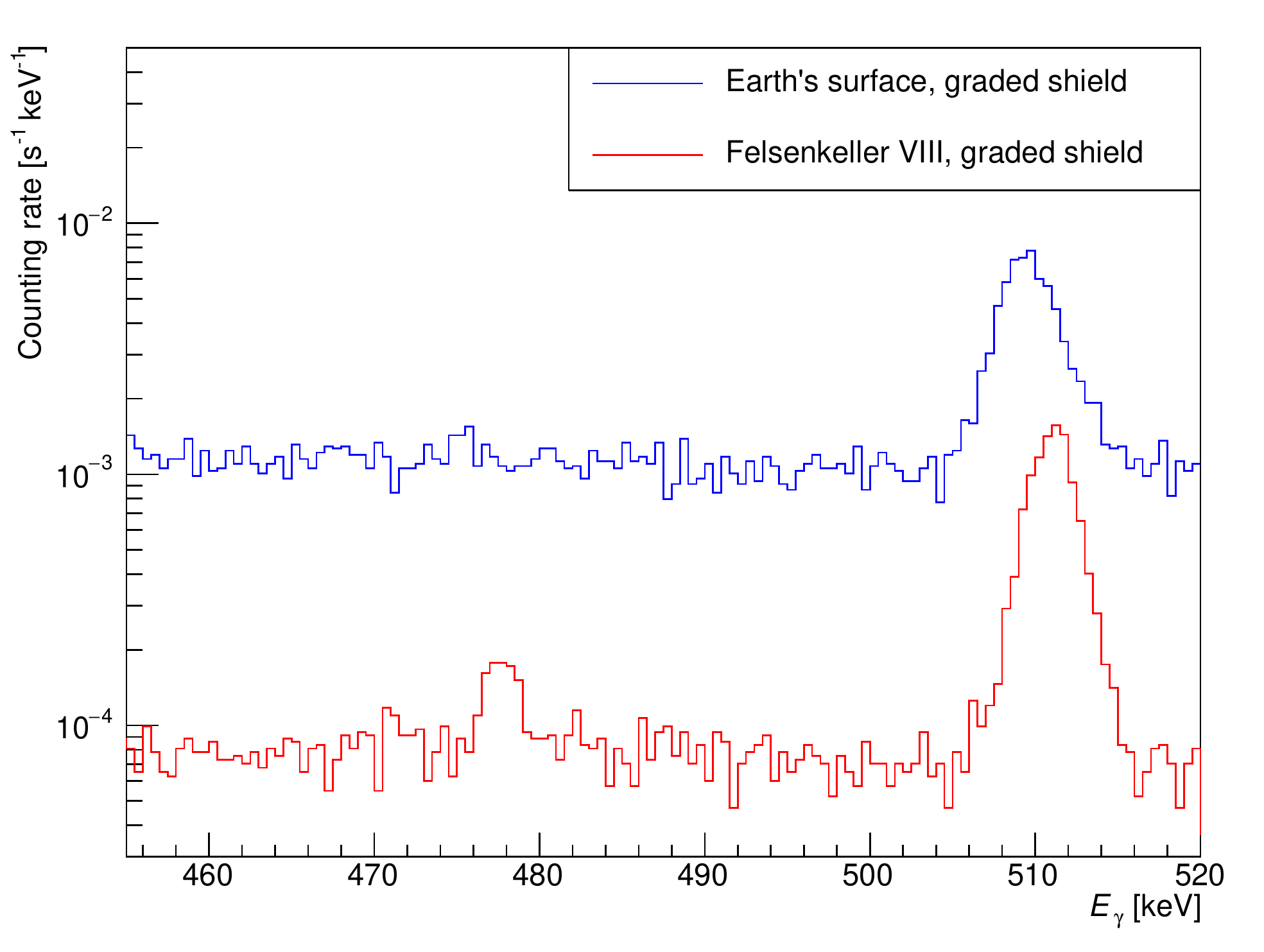}
\caption{One and the same activated $^7$Be sample from a recent study of the $^3$He($\alpha$,$\gamma$)$^7$Be reaction \cite{Huebinger18-BSc}, counted subsequently in a graded shield at the Earth's surface (HZDR Rossendorf \cite{Erhard09-Diss}) and underground in Felsenkeller (tunnel VIII), in a preliminary version of the new shielding. The $\gamma$ ray from the decay of the 478\,keV first excited state of $^7$Li to the ground state, characteristic of $^7$Be radioactivity, is only apparent in the underground spectrum.}
\label{fig:7Be-Felsenkeller}       % Give a unique label
\end{figure}
% =================================================================================

One particularly important example is the $^3$He($\alpha,\gamma$)$^7$Be reaction already discussed in Figure \ref{fig:3He4He}. For this reaction, there are three modern experiments with a systematic uncertainty as low as 3\% each \cite{NaraSingh04-PRL,Bemmerer06-PRL,Brown07-PRC} and a fourth one with just 5\% systematic uncertainty \cite{diLeva09-PRL} that are all essentially in agreement with the "Solar Fusion II" recommended fit curve \cite{Adelberger11-RMP}. Despite this favorable situation, the final evaluated error bar for the "Solar Fusion II" fit is 5\% because of the difficulty of extrapolation from the high-energy experiments \cite{NaraSingh04-PRL,diLeva09-PRL,Brown07-PRC} to the unique low-energy experiment at LUNA \cite{Bemmerer06-PRL}.

In order to push for higher precision, the low-energy $^3$He($\alpha,\gamma$)$^7$Be data must be connected by one and the same experiment spanning an energy range overlapping both with the low-energy, LUNA, and the high-energy, surface-based, data. This is planned at Felsenkeller. 

As a preparatory phase for such a study, both the angular distribution of the emitted $\gamma$-rays and the $^7$Be activation yield are currently under study at TU Dresden. The first part of the irradiations have already been carried out at the surface-based HZDR 3\,MV Tandetron accelerator in Dresden-Rossendorf. An example of a $^7$Be activation spectrum that was already obtained in the new facility in tunnel VIII is shown in Figure~\ref{fig:7Be-Felsenkeller}. It is clear from the Figure that only by underground $\gamma$-counting the necessary sensitivity for a wide energy range study of this fundamentally important cross section can be reached.

% =================================================================================
\section{Stellar helium burning and the $^{12}$C($\alpha,\gamma$)$^{16}$O reaction}
\label{sec:c12ag}

Stellar helium burning provides the third and fourth (after $^1$H and $^4$He) most abundant nuclides, $^{12}$C and $^{16}$O, which form the building blocks for all future nucleosynthetic processes. The ratio of the carbon and oxygen abundances has a deep effect on many other elemental abundances \cite{Weaver93-PhysRep}, and it is therefore necessary to fix its predicted value based on laboratory experiments.

The predicted $^{12}$C/$^{16}$O abundance ratio depends on the rates of two nuclear reactions at the temperatures of stellar helium burning, $T_9$ $\lesssim$ 0.35 (with $T_9$ the stellar temperature in GK):

\begin{eqnarray}\rm
^4He + ^4He \rightarrow ^8Be + ^4He & \rightarrow & \rm ^{12}C \label{eq:3alpha} \\
\rm ^{12}C + ^4He & \rightarrow & \rm ^{16}O + \gamma \label{eq:c12ag}
\end{eqnarray}
The rate of reaction (\ref{eq:3alpha}), the so-called triple-$\alpha$ reaction, depends strongly on the properties of the 0$^+$ excited state at 7654 keV (also known as the "Hoyle state") in $^{12}$C. In recent times, this state is even under study by {\it ab initio} theoretical methods \cite{Epelbaum12-PRL}, so it may be hoped that the triple-$\alpha$ rate will soon be precisely known. 

% =================================================================================
\begin{figure}[tb]
\centering
\includegraphics[width=\textwidth]{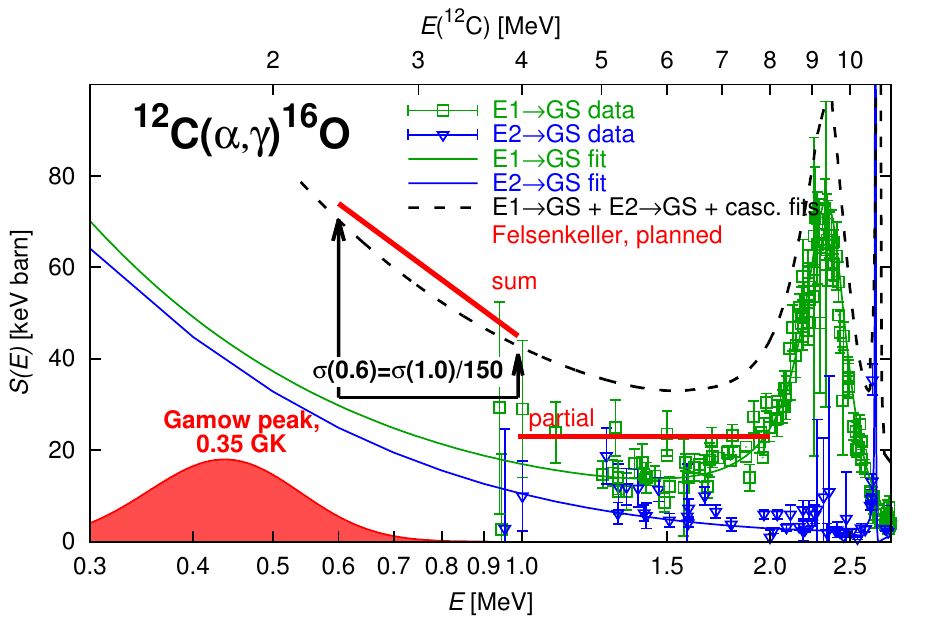}
\caption{Excitation function of the $^{12}$C($\alpha,\gamma$)$^{16}$O reaction. Experimental data \cite{NACRE13-NPA} and R-matrix fits for the dipole \cite{NACRE13-NPA}, quadrupole \cite{Brune15-ARNPS}, and sum of all com\-po\-nents \cite{NACRE13-NPA} are plotted. The feasible energy range for direct total and partial cross section data at Felsenkeller, based on the measured background \cite{Szucs12-EPJA,Szucs15-EPJA} and assuming inverse kinematics, and the Gamow peak for 0.35\,GK temperature are also shown.}
\label{fig:c12ag}       % Give a unique label
\end{figure}
% =================================================================================

However, the rate of reaction (\ref{eq:c12ag}), $^{12}$C($\alpha,\gamma$)$^{16}$O, is much more uncertain due to the complicated structure of the excitation function with a number of resonances near the Gamow window (Figure \ref{fig:c12ag}).

Direct cross section data reach down to barely below $E$ = 1\,MeV center of mass energy, and the energies of the Gamow peak are only accessible by R-matrix extrapolations \cite{deBoer17-RMP}, which carry uncertainties when they cannot be validated by experimental data over a wide energy range. 

At Felsenkeller, it is planned to solve the $^{12}$C($\alpha,\gamma$)$^{16}$O puzzle with direct experiments in inverse kinematics, bombarding a helium target with an accelerated $^{12}$C beam. Using the known (see below, section \ref{sec:labbg}) background in Felsenkeller, it can be shown that partial cross sections, requiring HPGe detectors and good angular resolution, can be obtained in the $E$ = 1.0-2.0\,MeV energy range. Using a large, high-efficiency $\gamma$-ray detector such as the ones used for total absorption spectroscopy \cite[e.g.]{Tain15-PRL}, an even lower energy range,  $E$ = 0.6-1.0\,MeV, can be covered (Figure \ref{fig:c12ag}). 

These planned new experimental data will enter the Gamow peak for helium burning for the first time and allow to put an experimental constraint that either confirms or refutes existing extrapolations.

% =================================================================================
\section{No-beam background studies}
\label{sec:labbg}

The feasibility of underground nuclear astrophysics experiments depends decisively on the remaining experimental background. 

\begin{figure}[b]
\centering
\includegraphics[width=\textwidth]{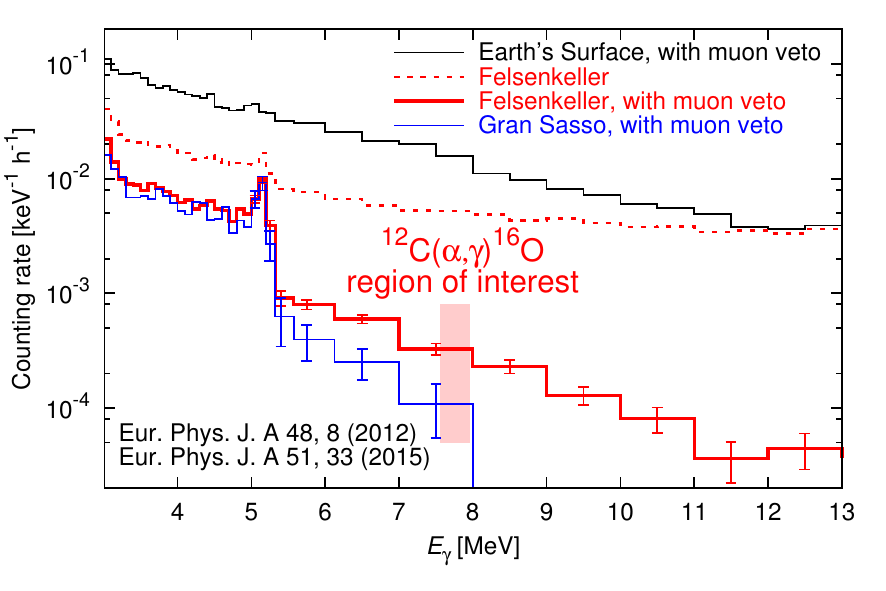}
\caption{Background intercomparison using one and the same escape-suppressed HPGe detector, subsequently transported to different sites. The data \cite{Szucs12-EPJA,Szucs15-EPJA} show that using a muon veto, the Felsenkeller background is not far from the deep underground (Gran Sasso) case.}
\label{fig:Szucs}       % Give a unique label
\end{figure}

In particular, the background in muon-vetoed $\gamma$-ray detectors was studied in details \cite{Szucs12-EPJA,Szucs15-EPJA}, and the data from these background studies were essential for the adoption of the project (Figure \ref{fig:Szucs}).

The laboratory background without ion beam in Felsenkeller depends on the remaining cosmic ray muon flux and on, mainly cosmic ray induced, neutrons. The former component can be partially suppressed by an active muon veto. Detailed studies carried out in Felsenkeller tunnel IV \cite{Szucs12-EPJA,Szucs15-EPJA} have shown that when using such an active veto, the background in the crucial 7-8 MeV energy window relevant for the study of the $^{12}$C($\alpha$,$\gamma$)$^{16}$O reaction is just a factor of three worse than deep underground in Gran Sasso, low enough that it does not limit most experiments (Figure \ref{fig:Szucs}). 

% =================================================================================
% =================================================================================

This background study has recently been repeated in Felsenkeller tunnel VIII, where the new accelerator lab is installed. The preliminary data show a background in muon vetoed $\gamma$ detectors that is consistent with that of tunnel IV. 

The root cause of the Felsenkeller background, the low but non-negligible remaining muon flux, has been studied in detail using a muon tomograph, which was deployed successively in tunnels IV, VIII, and IX, and which showed an almost constant integrated muon flux over those parts of the tunnel system which are furthest from the entrance (Felix Ludwig {\it et al.}, these proceedings).

Likewise, the (again low but non-negligible) flux and energy spectrum of neutrons in Felsenkeller, have also been studied in detail using $^3$He counters with and without moderators (Thomas Hensel {\it et al.}, these proceedings).

% =================================================================================
\section{Background studies with ion beam}
\label{sec:beambg}

\begin{figure}[b]
\centering
\includegraphics[width=\textwidth]{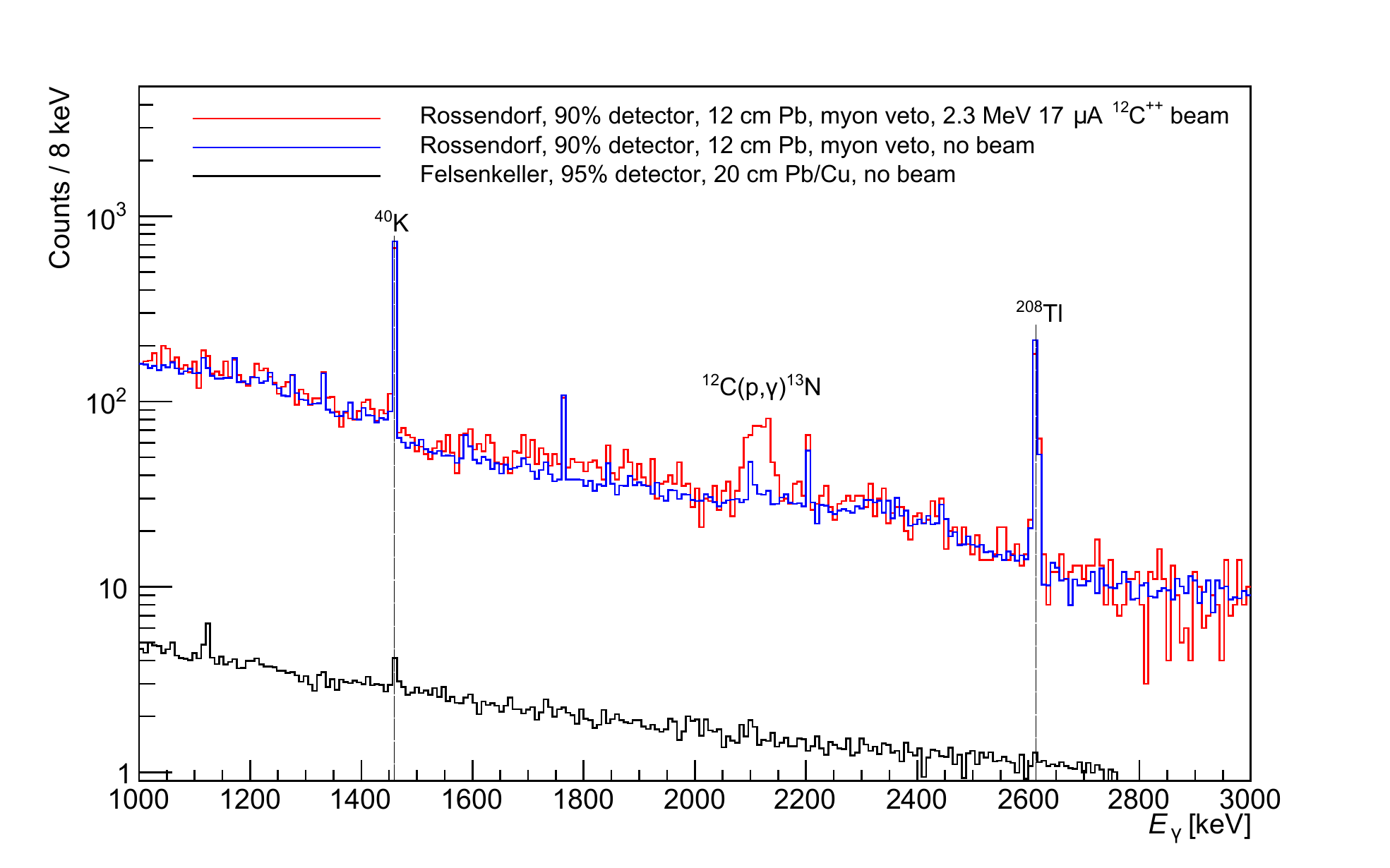}
\caption{Observed $\gamma$-ray spectrum with an actively and passively shielded, large HPGe detector in close geometry, with intensive $^{12}$C ion beam incident on a hydrated titanium target \cite{Reinicke18-Diss}.}
\label{fig:Reinicke}       % Give a unique label
\end{figure}

The low background conditions observed in underground experiments can only be fully exploited for accelerator-based studies if there is no additional background induced by the ion beam. As the Felsenkeller accelerator is not yet running, this point has been addressed in a recent experiment using an intensive, 9 particle-$\upmu$A, $^{12}$C beam at the surface-based HZDR 3\,MV Tandetron accelerator in Rossendorf (Figure \ref{fig:Reinicke}).

% =================================================================================
% =================================================================================

The $^{12}$C beam energy used for this test was 2.3\,MeV. This corresponds to a center-of-mass energy of $\sim$0.6\,MeV for an inverse-kinematics $^{12}$C($\alpha,\gamma$)$^{16}$O study, close to the lower limit of the energy range to be studied in the planned $^{12}$C($\alpha,\gamma$)$^{16}$O experiment (see above section \ref{sec:c12ag} and Figure \ref{fig:c12ag}). Due to the steep drop of the cross section to lower energies (see above section \ref{sec:intro} and eq. (\ref{eq:Sfactor})), the background situation is most important at the lowest beam energy, exactly the energy examined here. 

The beam intensity of the test is fivefold lower than what is projected for the Felsenkeller accelerator, 50 particle-$\upmu$A $^{12}$C beam, limiting the statistics. Furthermore, the statistics is limited by the high laboratory background observed at the Earth's surface, despite 12\,cm lead shielding and an additional active muon veto. 

Even still, it is apparent from Figure \ref{fig:Reinicke} that the continua of the two spectra with and without beam (red and blue curves, respectively, in Figure \ref{fig:Reinicke}) are consistent everywhere above the energy of the well-known $^{12}$C($p,\gamma$)$^{13}$N peak, indicating that there is no significant beam-induced $\gamma$-ray yield at 7-8 MeV, in the energy range of interest for the planned $^{12}$C($\alpha,\gamma$)$^{16}$O experiment.

These studies will be repeated in Felsenkeller, where due to the higher beam intensity and lower no-beam background a much higher sensitivity is expected, see the black curve in Figure \ref{fig:Reinicke}).

% =================================================================================
\section{The project for a new underground ion beam}
\label{sec:Project}

% =================================================================================
\begin{figure}[t]
\centering
\includegraphics[width=0.8\textwidth]{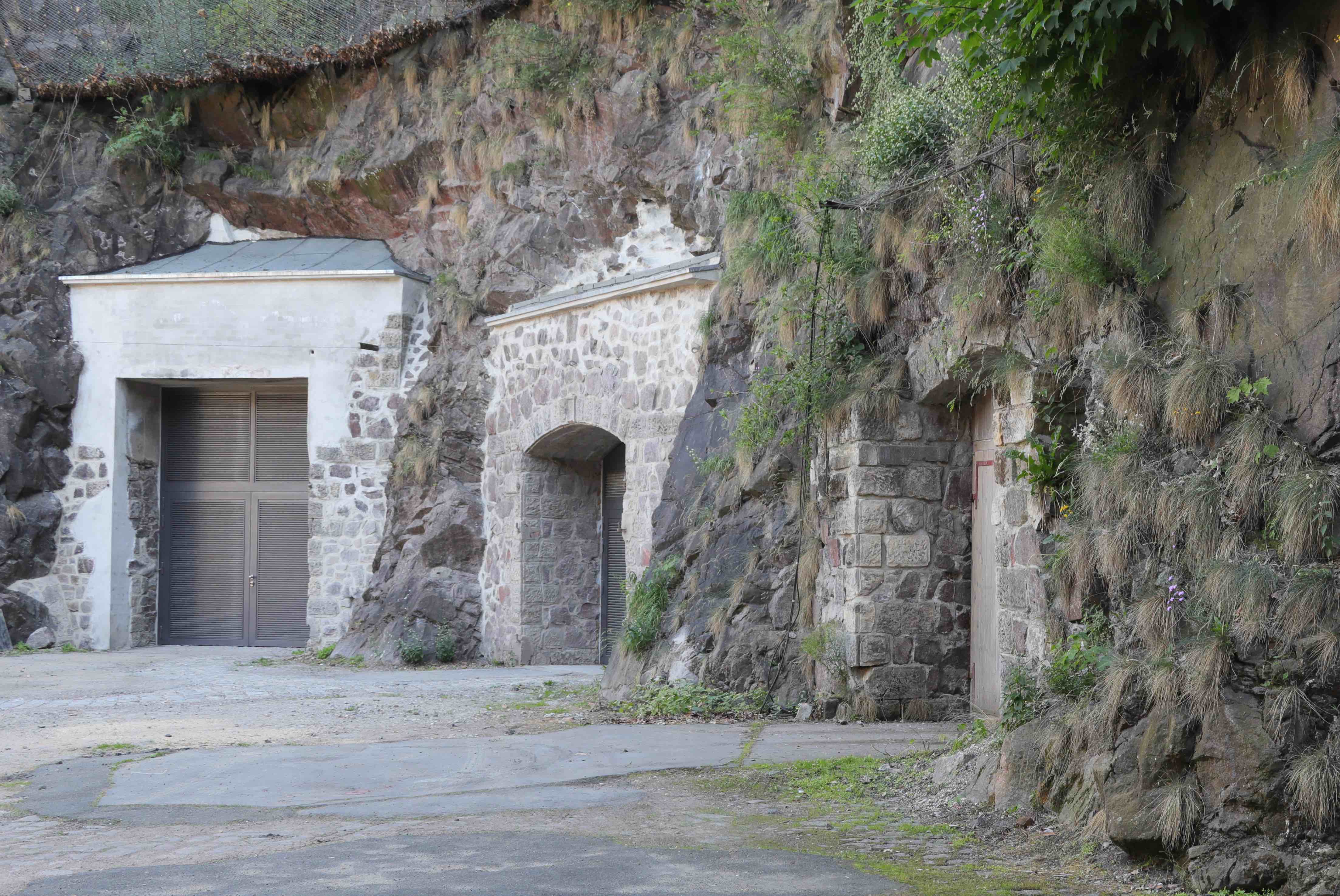}
\caption{Entrances of Felsenkeller tunnels (from left) IX, VIII, and VII. The new laboratory is situated in tunnels IX and VIII.}
\label{fig:Tunnelmund}       % Give a unique label
\end{figure}
% =================================================================================

In the years 1856-1859, nine tunnels, called tunnel I-IX, respectively, have been driven into the almost perpendicular hornblende monzonite rock wall of the Felsenkeller site in the Plauenscher Grund site in Dresden, immediately next to the river Weißeritz. At the time of construction, the tunnels were intended as cool storage for the Felsenkeller brewery, but over time they became open to other uses when technical refrigeration became commonplace (Figure \ref{fig:Tunnelmund}). 

Starting in 1982, tunnel IV is used for low-level $\gamma$-analytics \cite{Helbig84-Isotopenpraxis}, going on until today \cite{Koehler09-Apradiso}.

Based on this history, tunnels VIII and IX have recently been prepared for use in experimental nuclear astrophysics. Near the end of tunnel VIII, two measurement bunkers were erected, made of 40\,cm thick concrete with $\sim$20\,Bq/kg uranium and thorium content. Sand, gravel, and cement samples were studied by $\gamma$-spectrometry one day before each concrete was mixed and cast, in order to stop the casting process, if needed.

The two measurement bunkers and an adjacent accelerator hall are air-conditioned (and heated in winter) to keep 20\,$^\circ$C room temperature and the dewpoint below 15\,$^\circ$C. In addition to 4000\,m$^3$/h recirculated air, 500\,m$^3$/h fresh air are brought in from the outside of the tunnels. The laboratory space and the open tunnels are fitted with smoke detectors to detect fires and oxygen monitors to warn against release of suffocating gases, such as SF$_6$ (used for high voltage insulation), nitrogen, or argon. 

% =================================================================================
\begin{table}[tb]
\tbl{\label{Table:Sand} Top: Specific activities for the $^{232}$Th and $^{238}$U decay chains, assumed to be in equilibrium, from samples of the different components of the concrete used in Felsenkeller, casting \#3. Bottom: Weighted averages for the altogether three castings necessary for building the concrete bunkers and walls.}
{\resizebox{\textwidth}{!}{%
\begin{tabular}{|l|r|c|c|} \hline
\bf Component & Mass  & \multicolumn{2}{c|}{Specific activity [Bq/kg]} \\%[1mm] %\cline{3-4}
 & share & $^{232}$Th chain & $^{238}$U chain \\ \hline
Sand 0/2 mm Borsberg/Saxony 	& 34\% 	& 10.6$\pm$2.0 & 11.2$\pm$1.3 \\ 
Gravel 2/8 mm Borsberg/Saxony	& 10\%	& 25.9$\pm$3.6 & 20.3$\pm$2.7 \\
Gravel 8/16 mm Borsberg/Saxony	& 39\%	& 17.3$\pm$2.9 & 20.5$\pm$2.1 \\
Cement II/A-LL 32.5 R Cizkovice/Czech Rep.	& 16\%	& 19.1$\pm$4.4 & 15.9$\pm$2.6 \\
Fuel ash Opole/Poland	& 1.8\%	& 85$\pm$12 & 87$\pm$9 \\ \hline \hline

Weighted average, casting \#3 & & 17.4$\pm$3.1 & 17.8$\pm$2.1 \\ %\hline \hline
Weighted average, casting \#1 & & 16.3$\pm$2.1 & 14.9$\pm$1.4 \\ 
Weighted average, casting \#2 & & 15.6$\pm$1.9 & 17.0$\pm$1.7 \\ \hline %\hline
\end{tabular}%
}
}
\end{table}
% =================================================================================

There are two main installations in tunnels VIII and IX: First, a 5\,MV Pelletron accelerator, type 15SDH-2, made by National Electrostatics. This is a second-hand machine that was in use from 1999-2012 in York/UK for $^{14}$C analyses by accelerator mass spectrometry. It includes double charging chains, enabling 300\,$\upmu$A upcharge current, and an MC-SNICS 134 sputter ion source that is optimized for high output carbon beams. This source has since been tested successfully at low acceleration voltages \cite{Koppitz17-BSc}.

After the machine was acquired by HZDR in 2012, an additional internal ion source of the radio frequency type was added, in order to boost performance for $^1$H and $^4$He beams. The RF ion source is placed inside the high voltage terminal at an angle of 30$^\circ$, and the beam is bent by an electrostatic deflector into the axis of the accelerator \cite{Reinicke18-Diss}. The tests of the two ion sources are described elsewhere (Marcel Grieger {\it et al.}, these proceedings).

The second main installation is a 150\% ultra-low-background high-purity germanium (HPGe) detector for offline activity determination. It is hosted in a separate bunker and will be used for example for the activation analysis of the $^3$He($\alpha$,$\gamma$)$^7$Be cross section. The same detector can also be used and for study of ultra-low activity parts to be deployed in neutrino-less double beta decay studies \cite{Legend17-AIPCP}.

% =================================================================================
\begin{figure}[t]
\centering
\includegraphics[width=\textwidth]{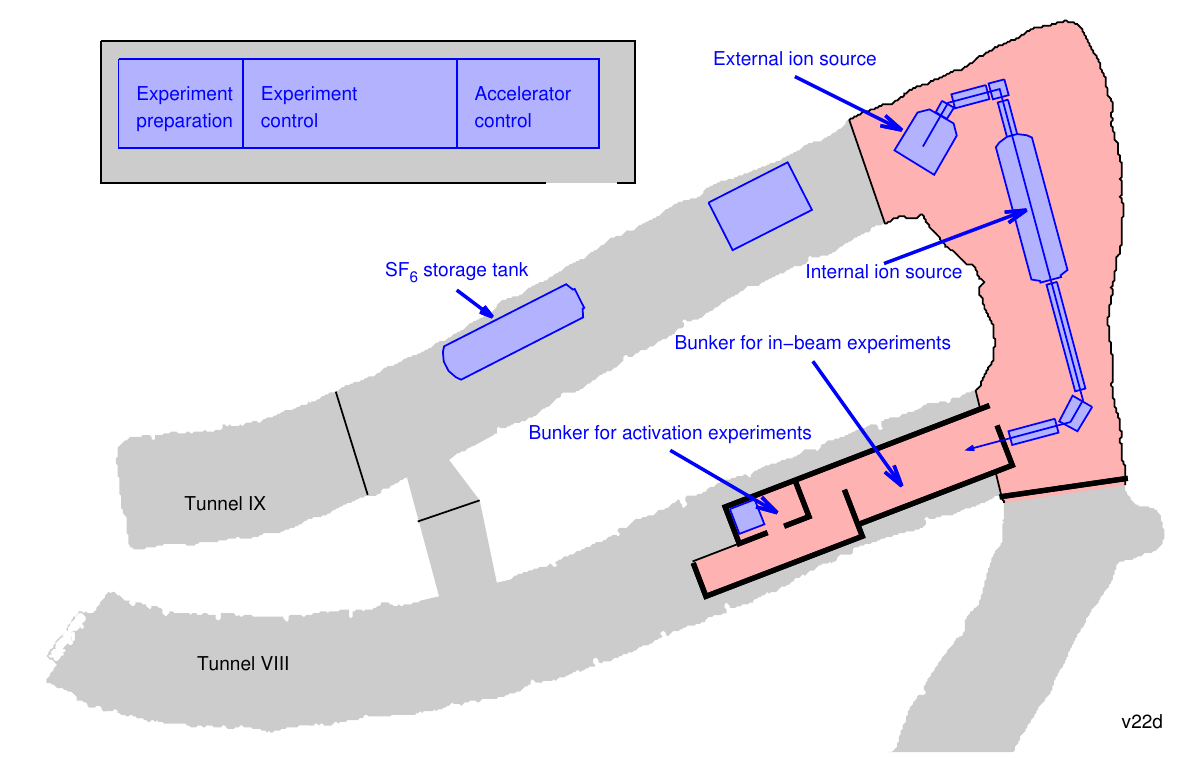}
\caption{Sketch of the most important installations in Felsenkeller tunnels VIII and IX. The thick black lines correspond to 40\,cm thick low-activity concrete walls.}
\label{fig:Map}       % Give a unique label
\end{figure}
% =================================================================================

% =================================================================================
\section{Status of the project}
\label{sec:Status}

% =================================================================================
\begin{figure}[t]
\centering
\includegraphics[width=\textwidth]{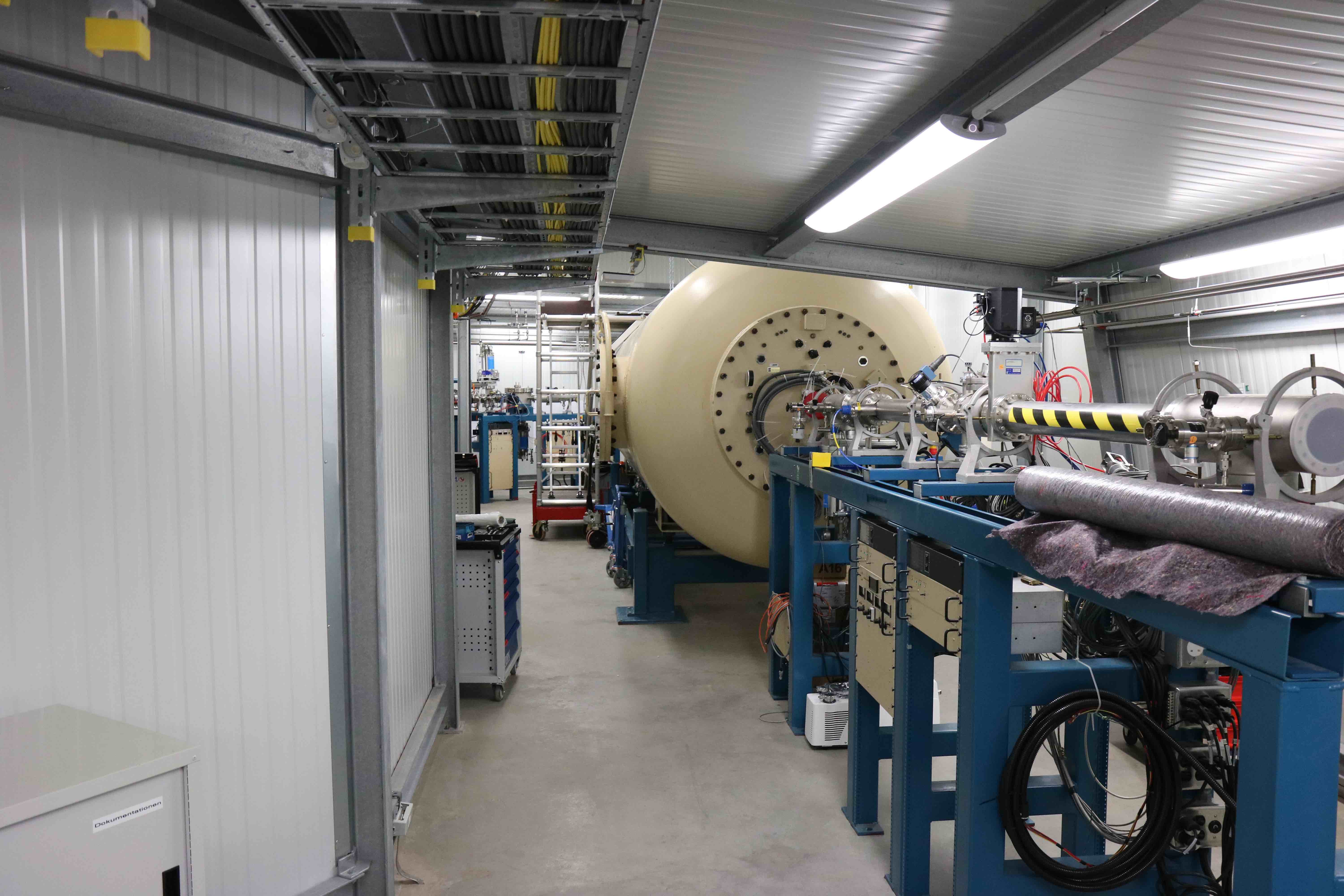}
\caption{View of the accelerator room underground, including the high voltage tank of the 5\,MV Pelletron. The beam lines are already placed in the tunnel, but not yet aligned and connected.}
\label{fig:AcceleratorRoom}       % Give a unique label
\end{figure}
% =================================================================================

All the main installations have already been transported underground. The 5\,MV Pelletron has been placed in its final position (Figure \ref{fig:AcceleratorRoom}). The beam transport system has already been put in place but still needs to be aligned. 

Likewise, the 150\% HPGe detector is underground and operational, but its shielding made of copper and low-background lead of typically $\lesssim$3 (for the inner layers), $\lesssim$30, and $\lesssim$50\,Bq/kg $^{210}$Pb is still being set up. The detector will also include an active veto made by four 1\,m$^2$ large, thick plastic scintillator paddles to gate out muon-induced signals.

The necessary permissions for operating the accelerator and for accessing and using the underground tunnels have already been issued by the competent authorities and are held by HZDR, which is responsible for day to day operations including all safety and security aspects. 

An interlock system monitoring accelerator, radiation dose, and laboratory safety sensors allows round-the-clock operation of the ion beam also without the presence of personnel on site. However, daily on-site checks by trained personnel are required, and likewise any change of accelerator parameters will require the presence of a trained accelerator operator on site.

% =================================================================================
\section{Outlook}
\label{sec:Outlook}

After completion of the commissioning of the underground installations, two in-house experiments are planned: 
\begin{enumerate}
\item First, a study of the solar-fusion $^3$He($\alpha$,$\gamma$)$^7$Be reaction using the (internal) terminal $^4$He$^+$ ion source, single-ended accelerator operation, and a $^3$He implanted or gaseous target. 
\item Second, a study of the $^{12}$C($\alpha$,$\gamma$)$^{16}$O reaction of helium burning in inverse kinematics, using the (external) sputter $^{12}$C$^-$ ion source, tandem accelerator operation, and a $^4$He implanted or gaseous target.
\end{enumerate}

The precise arrangement and sequence of these two "day one" experiments will depend on the level of readiness of the two respective ion sources needed. Since the target arrangement will be similar, and in both cases a number of well-shielded HPGe detectors will be needed, it will be possible to perform the experiments in close temporal sequence.

At the same time, a significant amount of beam time, of the order of 500 hours per year, will be available to outside users free of charge. Beam time proposals will be evaluated solely on their scientific merit by a panel of independent outside advisers, and full local support will be given for groups wishing to use the underground accelerator for any valid field of science, starting from but not restricted to nuclear astrophysics. 

It is hoped that this open access model will contribute to inter-comparability and overall accelerate progress in the exciting field of nuclear astrophysics, including the nuclear physics of the Sun.

\section*{Acknowledgments}

The authors are indebted to all the participants in the Fifth International Solar Neutrino Conference for a highly encouraging meeting with many new ideas and challenges. --- Financial support by DFG (TU Dresden Institutional Strategy "support the best", INST 269/631-1 FUGG, BE4100/4-1, and ZU123/21-1) and by the Helmholtz Association (NAVI HGF VH-VI-417 and ERC-RA-0016) is gratefully acknowledged.

% =================================================================================

\end{document}